\documentclass[review]{elsarticle}

\usepackage{hyperref}
\usepackage{epspdfconversion}
\journal{Journal of \LaTeX\ Templates}

\begin{document}

\begin{frontmatter}

\title{\textbf{Metamagnetism stabilized giant magnetoelectric coupling in ferroelectric \textit{x}}BaTiO${_3}$-(1-\textit{x})BiCoO${_3}$ solid solution}

\author[add1,add2]{Lokanath Patra}
\author[add3,add4]{Zhao Pan}
\author[add3]{Jun Chen}
\author[add5]{Masaki Azuma}
\author[add1,add2,add6,add7]{P. Ravindran}
\ead{raviphy@cutn.ac.in}

\address[add1]{Department of Physics, Central University of Tamil Nadu, Thiruvarur 610101, India}
\address[add2]{Simulation Center for Atomic and Nanoscale MATerials, Central University of Tamil Nadu, Thiruvarur, Tamil Nadu, 610101, India}
\address[add3]{Department of Physical Chemistry, University of Science and Technology Beijing, Beijing 100083, China}
\address[add4]{State key laboratory of refractories and metallurgy, Wuhan University of Science and Technology, Wuhan 430081, China}
\address[add5]{Materials and Structures Laboratory, Tokyo Institute of Technology, 4259 Nagatsuta, Midori, Yokohama 226-8503, Japan}
\address[add6]{Department of Materials Science, Central University of Tamil Nadu, Thiruvarur, Tamil Nadu, 610101, India}
\address[add7]{Center for Materials Science and Nanotechnology and Department of Chemistry, University of Oslo, Box 1033 Blindern, N-0315 Oslo, Norway}
\begin{abstract}
In order to establish the correlation between the magnetoelectric coupling and magnetic instability, we have studied the structural, magnetic, and ferroelectric properties of BaTiO$_3$ modified BiCoO$_3$ i.e. \textit{x}BaTiO${_3}$-(1-\textit{x})BiCoO${_3}$ as a function of BaTiO$_3$ concentration ($x$) and volume from a series of general$-$gradient$-$ corrected, full potential, spin$-$density$-$functional band$-$structure calculations within the framework of density functional theory along with synchrotron X$-$ray diffraction and magnetic measurement studies .Our total energy calculation shows that the $G-$type antiferromagnetic (AFM) spin configurations with both inter$-$ and intra$-$layer AFM coupling were found to be energetically favorable among all the considered magnetic configurations for $x<$ 0.45 and higher concentrations stabilize with nonmagnetic (NM) states. We observe metamagnetic spin state transitions associated with paraelectric to ferrolectric transitions as a function of volume and $x$ using  synchrotron diffraction and computational studies, indicating a strong magnetoelectric coupling. Specifically for $x=$ 0.33 composition, a pressure induced high spin (HS) to low spin (LS) transition occurs when the volume is compressed below 5\%. Our orbital$-$projected density of states show a HS state for Co$^{3+}$ in the ferroelectric ground state for $x<$0.45 and the corresponding paraelectric phase is stable in the NM state due to the stabilization of LS state as evident from our fixed$-$spin$-$moment calculations and magnetic measurements. Chemical bonding has been analyzed using partial density of states, electron localization function(ELF), and Born effective charge analysis. Our ELF analysis confirms the presence of stereochemically active lone pair electrons at the Bi sites which are mainly  responsible for the off$-$center displacements of Bi atoms. High values of spontaneous ferroelectric polarizations are predicted for lower $x$ values which inversely vary with $x$ because of the reduction of tetragonality ($c/a$) with increase in $x$ which indicates the presence of both spin-lattice and ferroelectricity-lattice. Our partial polarization analysis shows that not only the lone pair at Bi sites but also the $d^0-$ness of Ti$^{4+}$ ions contribute to the net polarization. Moreover, we find that the HS$-$LS transition point and magnetoelectric coupling strength can be varied by $x$.
\end{abstract}


\end{frontmatter}

\section{Introduction}

In a broad range of industrial and scientific fields, the research of perovskite materials (ABO$_3$) with both high Curie temperature (T$_C$) and excellent ferroelectric(magnetic) performance are widespread because of their application in sensors, actuators, transducers, and so on within a wide temperature range.\cite{park2002high,scott2007applications}  Magnetoelectric (ME) multiferroics with coupled (anti$-$)ferroelectric and (anti$-$)ferromagnetic ordering have gained attention as they provide the opportunity for developing tunable devices.\cite{ramesh2007multiferroics,cheong2007multiferroics,eerenstein2006multiferroic,fiebig2005revival,nan2008multiferroic} Bi$-$based multiferroics have become a center of attraction due to the high value of electric polarization,\cite{ravindran2006theoretical,ravindran2008magnetic} and Bi$^{3+}$ is more environment$-$friendly than Pb$^{2+}$. In addition, the stereochemical activity of Bi$^{3+}$ is being exploited in magnetic oxides, with the goal of forming the ferromagnetic-ferroelectric coupling. Recently, Bi$-$based perovskites of BiMeO$_3$, in which Me are cations with an average valence of +3, have been adopted to form a binary system with BaTiO$_3$, so that the high$-$temperature ferroelectric properties are optimized.\cite{suchomel2005enhanced,choi2005structure,chen2006structure} Furthermore, the $B-$site cations also introduce new properties. BaTiO$_3-$based perovskite compounds are significant multifunctional materials, which have been extensively researched in the last half century.\cite{haertling1999ferroelectric,cooper2012driven,chen2013unusual}Through chemical modification using BaTiO$_3$, we can manipulate the physical properties of many compounds like ferroelectric, piezoelectric, and other properties.\cite{grinberg2007structure}  For example, investigations on the multiferroic material are particularly of interest in BaTiO$_3-$BiFeO$_3$ (BF$-$BT).\cite{zhu2008structural,cotica2012high} This series exhibits excellent ME properties and high T$_C$ (T$_C$ of 580$-$6190$^0$C).\cite{leontsev2009dielectric} Kumar \textit{et al.}\cite{mahesh1998dielectric} reported that BF$-$BT system changes from rhombohedral to cubic symmetries for 0.1 $<x<$ 0.7, and then to a tetragonal symmetry for $x<$ 0.1. However, a non$-$centrosymmetric tetragonal symmetry was reported by Kim \textit{et al.}\cite{kim2004weak}  by neutron diffraction studies for $x<$ 0.6. In fact, the substitution of Ba$^{2+}$ for Bi$^{3+}$ and Ti$^{4+}$ for Co$^{3+}$ lead to more complex structural changes. 
\par
It is also worth noting that in BF$-$BT solid solutions two competing mechanisms for ferroelectricity coexist, i.e., the lone pair of the Bi atoms that shifts the Fe/Ti ion and also the Ti$-$O covalent bond that brings off$-$center displacements. In fact, these are the individual mechanisms for ferroelectricity existence in BF and BT compounds, respectively. However, (1-$x$)BiFeO$_3-x$BaTiO$_3$ system has some drawbacks. Firstly, the coexistence of Fe$^{2+}$ and Fe$^{3+}$ in these ceramics results in too high dc electrical conductivity at high temperatures due to a phonon$-$assisted electron hopping mechanism.\cite{zhou2012remarkably}  Therefore, the system must be sintered in an oxygen atmosphere and needs chemical modification to improve its resistivity.\cite{leontsev2009dielectric} Secondly, chemical doping has been strategically employed to improve the electrical properties in BiMeO$_3-$based ceramics. Nevertheless, most of the dopants added or substituted into the (1-$x$)BiFeO$_3-x$BaTiO$_3$ system lattice tend to lower T$_C$, limiting the temperature range of its use.\cite{zhou2012remarkably,fujii2011structural}
\par
It may be noted that BiCoO$_3$ is isostructural with tetragonal BaTiO$_3$ (space group $P4mm$) where the Co$^{3+}$ (3$d^6$) ion adopts high$-$spin configurations at ambient conditions with a magnetic moment of 2.75$\mu_B$/Co. The Co$^{3+}$ atoms go from high spin (HS) to low spin (LS) state around 5\% volume compression indicating the presence of metamagnetism in the system. Isolated layers of corner$-$shared CoO$_5$ pyramids are formed due to strong tetragonal distortion and the presence of lone pair electrons from the Bi$^{3+}$ ion brings offcenter displacement resulting ferroelectricity. The tetragonality ($c/a$ = 1.27, where $a$ and $c$ are the lattice constants) of BiCoO$_3$ is much larger than that of BaTiO$_3$, which indicates a larger ferroelectric polarization ($P_s$).\cite{belik2006neutron,oka2010pressure} Our previous first$-$principles Berry$-$phase calculations evaluated a giant electric polarization of 179$\mu$Ccm$^{-2}$ for BiCoO$_3$. In addition, the $C-$type antiferromagnetic ($C-$AFM) configuration was found to be the most stable phase among all the possible spin configurations considered for our total energy calculations. It was confirmed by the neutron powder diffraction (NPD) experiment that BiCoO$_3$ is a $C-$AFM insulator with the N\'eel temperature (T$_N$) of 470\,K.\cite{belik2006neutron} It is interesting to note that the antiferromagnetic ferroelectric ground state tetragonal $P4mm$ structure changes to a non$-$magnetic cubic paraelectric \textit{Pm-3m} structure under volume compression. So, we have concluded that BiCoO$_3$ is having strong coupling between magnetic and electric order parameters. This giant electromagnetic coupling is resulting from metamagnetism originating from HS$-$LS state transition of Co$^{3+}$ ion. So, tetragonal phase of BiCoO$_3$ is a promising candidate to use as ME material for various applications. However, difficulty in synthesizing this compound in large quantity(this material is usually synthesized by high pressure high temperature technique) hamper its use. This motivated us to look for ME behavior in composite MEs based on BiCoO$_3$. 
\par
One can tune the properties of complex transition metal oxides by cation substitutions, where the $B-$ site could exhibit fascinating cooperative electric order parameters i.e. spin, orbital and electron order. The coupling between lattice and these order parameters could bring out novel physical phenomena, in which lattice is coupled with ferroelectric, ferromagnetic, and orbital degrees of freedom resulting colossal magnetoresistance\cite{fang2000phase}, high T$_C$ superconductivity\cite{carbone2008direct}, multiferroism\cite{son2013four} etc. The flexibility in the crystal structure of bismuth$-$based perovskites provides an opportunity to explore unusual physical properties by chemical modification at $A$ and/or $B-$sites. Hence, in this work, the structural and the ME properties of composite ferrolectric $x$BaTiO$_3-$(1-$x$)BiCoO$_{3}$ were investigated using \textit{ab-initio} total energy calculations.  On one hand, these compounds are attractive due to the absence of toxic Pb content and on the other hand, they enhance physical properties compared with conventional ferroelectric materials. Further, as we have predicted giant ME coupling in BiCoO$_3$, the coupling can be tuned by varying the $x$ and through that identify potential new materials with desired ME coupling strength. This motivated us to take the present study.

\section{Experimental and Computational Details}
\subsection{Sample preparation}
All samples were prepared with a cubic anvil-type high-pressure apparatus. Stoichiometric mixture powder of BaO, TiO$_{2}$, Bi$_{2}$O$_{3}$, and Co$_{3}$O$_{4}$ was sealed in a gold capsule and reacted at 6\,GPa and 1473\,K for 30 min. A 10\,mg amount of the oxidizing agent KClO$_{4}$ (about 10 wt\% of the sample) was added to the top and bottom of the capsule in a separate manner. The obtained sample was crushed and washed with distilled water to remove the remaining KCl. After high pressure synthesis, the samples were carefully grounded and annealed at 673\,K for 1 hour and slowly cooled to room temperature. The X-ray diffraction (XRD) patterns were collected with the Bruker D8 ADVANCE diffractometer for phase identification at Argonne National Laboratory.
\subsection{Magnetic measurement}
The temperature dependence of the magnetic susceptibility (ZFC/FC) was measured with a SQUID magnetometer (Quantum Design, MPMS XL) in an external magnetic field of 1000 Oe.
\subsection{First-principle calculations}
First$-$principles DFT calculations were performed using the Vienna \textit{ab-initio} simulation package VASP,\cite{kresse1996software} within the projector augmented wave (PAW) method\cite{blochl1994projector} as implemented by Kresse and Joubert.\cite{kresse1996software}  In this approach, the valence orbitals are expanded as plane waves and the interactions between the core and valence electrons are described by pseudopotentials. Since one of the main aims of the present work is to determine ground-state structures for \textit{x}BatiO$_{3}-$(1-\textit{x})BiCoO${}_{3}$, we have performed structural optimization using force and stress minimization. The optimization of the atomic geometry is performed via a conjugate-gradient minimization of the total energy, using Hellmann-Feynman forces on the atoms and stresses in the unit cell. During the simulations, atomic coordinates and axial ratios are allowed to relax for different volumes of the unit cell. These parameters are changed iteratively so that the sum of lattice energy and electronic free energy converges to a minimum value. Convergence minimum with respect to atomic shifts is assumed to have been attained when the energy difference between two successive iterations is less than 10$^{-6}$\,eV per cell and the forces acting on the atoms are less than 1\,meV{\AA}$^{-1}$. Among the considered various crystal structures, the structure with the lowest total energy is taken as the ground- state structure.
\par
 The generalized gradient approximation (GGA)\cite{perdew1996generalized} includes the effects of local gradients of the charge density and generally gives better equilibrium structural parameters than the local density approximation (LDA). Ferroelectric properties are extremely sensitive to structural parameters and atom positions and hence we have used GGA for all our studies. The present study we have used GGA with the Perdew$-$Burke$-$Ernzerhof (PBE)\cite{ernzerhof1999assessment} and this functional gave good structural parameter for close relevant system BiFeO3.\cite{ravindran2006theoretical} We have used 800\,eV plane$-$wave energy cutoff for all the calculations. The calculations are carried out using 6x6x6 Monkhorst$-$Pack \textbf{k}$-$point mesh~\cite{monkhorst1976special} centered at the irreducible Brillouin zone for the ferroelectric \textit{P4mm}. We have used same energy cutoff and \textbf{k}$-$point density for all the calculations. The computations were performed in paramagnetic(PM), ferromagnetic(FM) and three antiferromagnetic (AFM) i.e. \textit{A}$-$type antiferromagnetic (\textit{A}$-$AFM; where the intraplane exchange interaction is ferromagnetic and interplane interaction is antiferromagnetic), \textit{C}$-$type antiferromagnetic (\textit{C}$-$AFM; where the intraplane exchange interaction is antiferromagnetic and interplane interaction is ferromagnetic), and $G-$type antiferromagnetic (\textit{G}$-$AFM; where both intraplane and interplane exchange interactions are antiferromagnetic) configurations.\cite{patra2016electronic} The Born effective charge are calculated using the so$-$called ``Berry phase finite difference approach'' in which the electronic contribution to the change of polarization is estimated using the modern theory of polarization.\cite{resta1994macroscopic,king1993theory} For the \textbf{k}$-$space integrations in the Berry$-$phase calculations, a uniform 8x8x8 \textbf{k}$-$point mesh was found to be adequate for BiCoO$_{3}$ and the same density of \textbf{k}$-$point is used for all the other compositions. In order to study metamagnetism, it is advantageous to calculate the total energy as a function of the magnetic moment using the so$-$called fixed$-$spin$-$moment method,\cite{schwarz1984itinerant} where one uses the magnetic moment \textit{M} as an external parameter and calculates the total energy as a function of \textit{M}. So, we have adopted the same approach to investigate the metamagnetism in the present study.
\section{Results and discussions}
\subsection{Structural details}
BiCoO$_{3}$ has a tetragonal perovskite structure with space group $P4mm$\cite{belik2006neutron}  so as tetragonal BaTiO$_{3}$\cite{smith2008crystal}  but with more electrical polarization.\cite{ravindran2008magnetic} In our previous work, pressure driven spin transition was observed in BiCoO$_{3}$ where the ground state ferroelectric $P4mm$ state changes to paraelectric $Pm-3m$ state around 5\% volume compression. Also, the paraelectric phase resembles a non-magnetic LS state where the ferroelectric phase shows a magnetic HS state showing the coupling between the magnetic and electric order parameter is very high.  Giant ME coupling was also found in BiCoO$_{3}$ which is originating from the HS$-$LS transition of Co$^{3+}$ ions.\cite{ravindran2008magnetic} We have substituted BaTiO$_{3}$ in BiCoO$_{3}$ to get composite ferroelectrics with composition $x$BaTiO$_{3}-$(1-$x$)BiCoO$_{3}$ where the $A$ site is occupied by Bi and Ba and B site is occupied by Co and Ti. We have performed first$-$principle calculations for $x$ = 0.25, 0.33, 0.5, 0.67, 0.75 and 1 compositions. To understand the pressure dependence of the lattice parameters and spin state transition for 0.33BTBCO, we have performed the total energy calculation as a function of volume for the fully relaxed experimental structure.  In cobaltites, the HS to LS transition can be obtained by doping,\cite{ravindran2002itinerant} temperature driven route, and pressure driven route.\cite{ravindran2008magnetic}
\par
\begin{figure}[!t]
\centering
\includegraphics[scale=0.1]{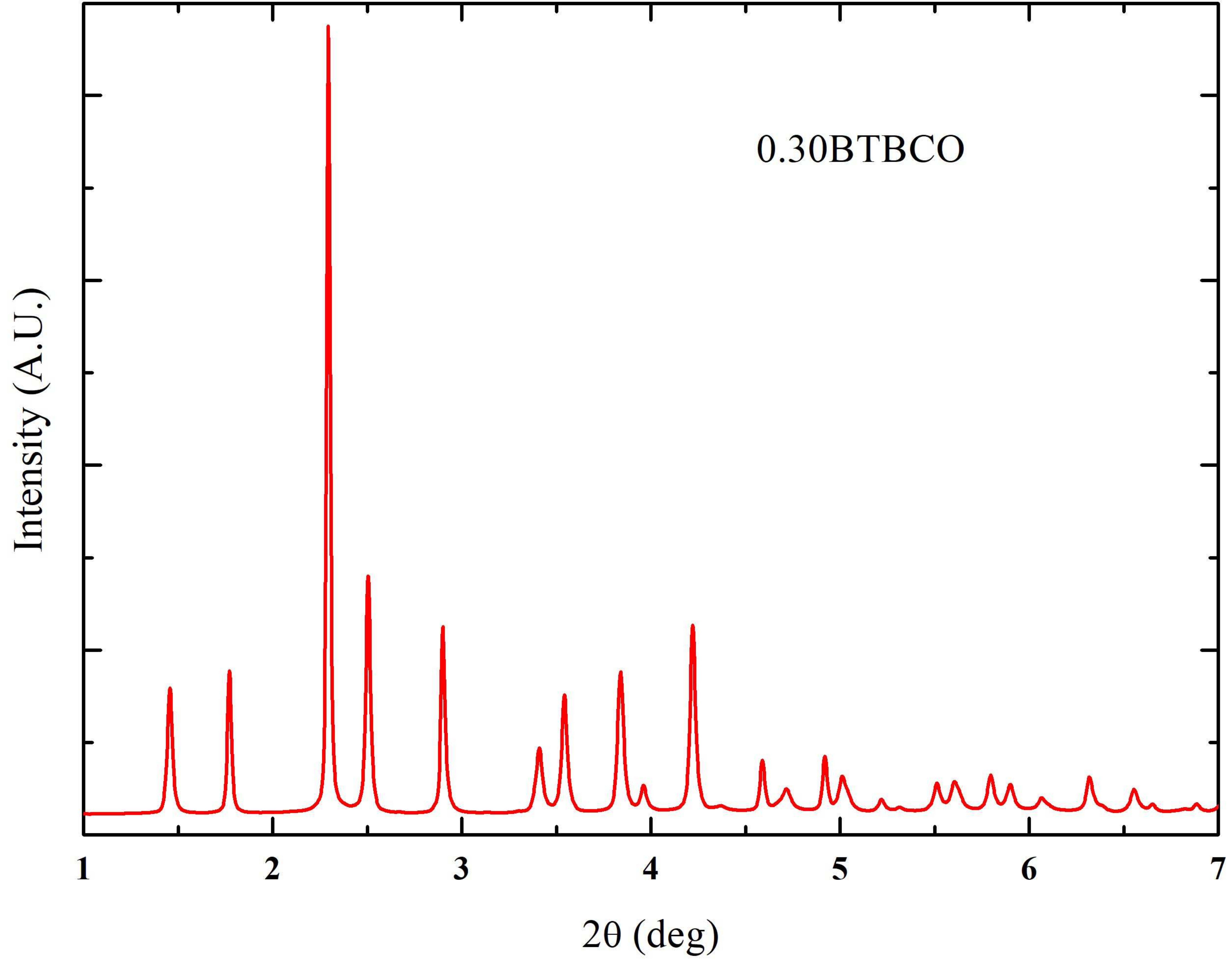}
\caption{\small(Color online) (Color online) X$-$ray powder diffraction patterns for 0.30BTBCO refined with the tetragonal structure (space group P4mm) at 300 K.}
\label{fig1}
\end{figure}
\begin{figure}[!t]
\centering
\includegraphics[scale=0.3]{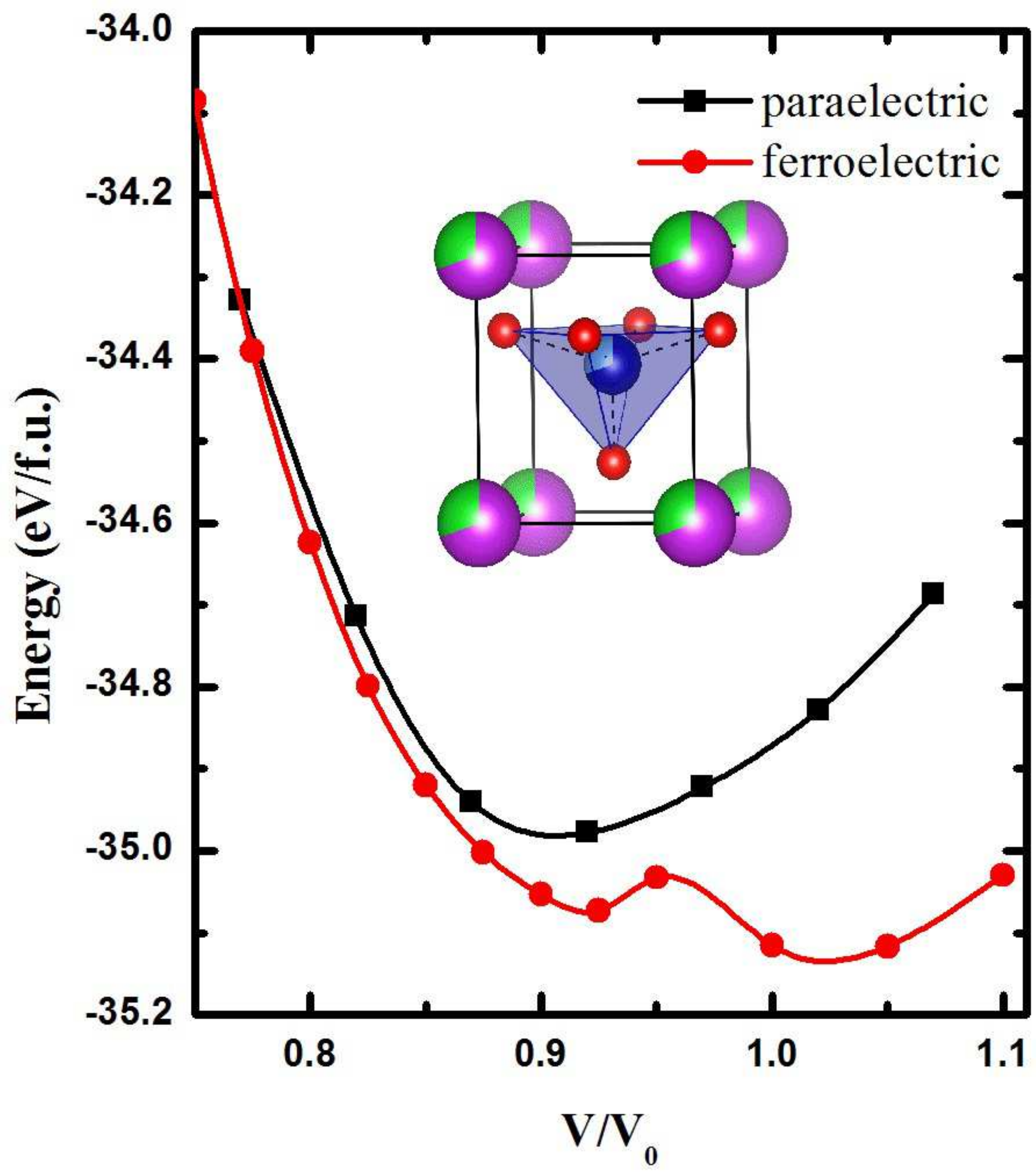}
\caption{\small(Color online) 
The crystal structure and total energy vs volume curves for ferroelectric and paraelectric phases of 0.33BTBCO. It can be seen that Ba(Bi) and Ti(Co) has 33\%(67\%) occupancy at the $A$ (corners  of the crystal) and $B$ (inside the polyhedra) sites respectively. The red color(smallest) spheres are oxygens. It is to be noted that there are two types of oxygens present in the system, one at the apical (O1) and four at the base (O2).}
\label{fig2}
\end{figure}

Figure~\ref{fig1} shows the XRD patterns for 0.2BTBCO compounds which was collected by high-energy synchrotron X$-$ray diffraction (SXRD) of wave length 0.117\,\AA. The structure refinement of $x$BaTiO3$-$(1-$x$)BiCoO$_3$ was done with the starting structural model (space group $P4mm$) which is based on the crystal structure of BaTiO$_3$ where the atomic positions of Ba/Bi is fixed at (0, 0, 0), Ti/Co at (0.5, 0.5, $z_{Ti/Co}$), O1 at (0.5, 0.5, $z_{O1}$) and O2 at (0.5, 0, $z_{O2}$).  The SXRD pattern for 0.2BTBCO was fitted well by this model. The SXRD pattern for other compositions are given in the supporting information.Our experimental crystal structures data obtained from  SXRD measurements were optimized using VASP. The calculated total energy versus volume curves for ferroelectric and paraelectric phases of 0.33BTBCO are given in Figure~\ref{fig2}. The equilibrium volume obtained from Figure~\ref{fig2} using the Birch$-$Murnaghan equation of states (EOS)\cite{birch1947finite}  and the corresponding calculated lattice parameters are in good agreement with our corresponding experimental values. The bulk modulus and its pressure derivative are found to be 105 GPa and 26.6, respectively for 0.33BTBCO. The calculated total energy curves for ferroelectric phase show two minima and one saddle point. This kind of feature was already observed in BiCoO$_{3}$ and this was associated with the metamagnetism originating from spin state transition. Similar to BiCoO$_{3}$ a pressure$-$induced metamagnetic transition is observed in 0.33BTBCO where the Co$^{3+}$ transforms from magnetic HS(ferroelectric) state to non$-$magnetic low$-$spin(paraelectric) state which can be seen in the discontinuity in E vs. V curve shown in Figure~\ref{fig2}. It is interesting to note that the equilibrium volume for the paraelectric phase is closer to a volume corresponding to the local minimum in the ferroelectric phase. If we reduce the volume further ferroelectric and the paraelectric curves coincide and this indicates the presence of ferroelectric to paraelectric phase transition in the system.  A metamagnetic transition refers to nonmagnetic materials which change to (anti)ferromagnetic materials with application of a sufficiently large amount of magnetic field. These types of band magnetism are mainly observed in Co$-$based materials. Similar metamagnetic behavior is also observed in the total energy curves of LaCoO$_{3}$ previously.\cite{ravindran2002itinerant}
\par
\begin{figure}[!t]
\centering
\includegraphics[height=4cm]{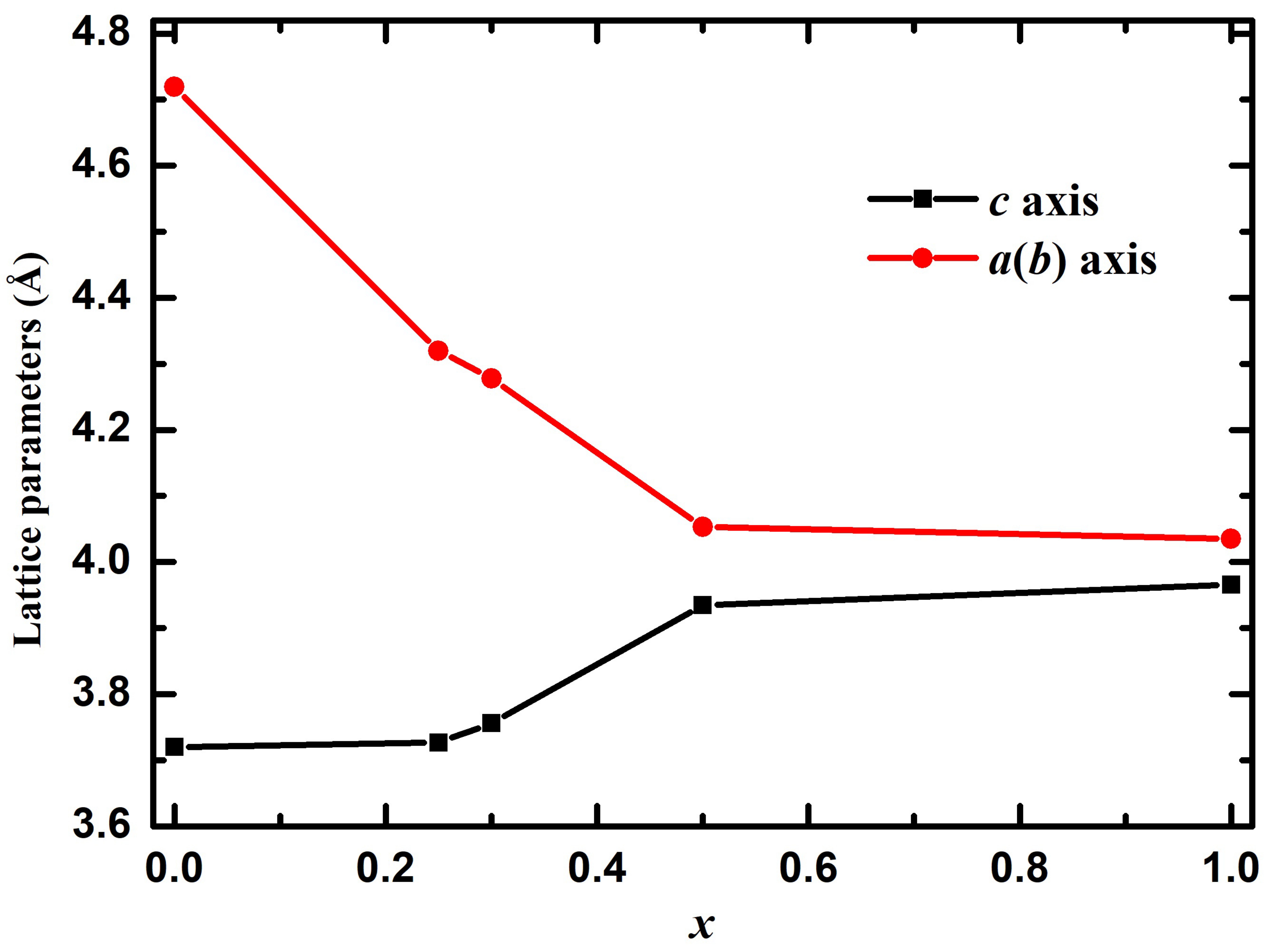}
   \hspace{0.10in}
    \includegraphics[height=4cm]{c_a-epspdf-to.pdf}
   \hspace{0.10in}
\caption{\small(Color online) (a) Evolution of lattice parameters of $x$BaTiO$_3-$(1-$x$)BiCoO$_{3}$ as a function of $x$. The rapidly decreasing $a-$axis and the slowly increasing $c-$axis make the tetragonality ($c/a$) to decrease continuously with the substitution of BaTiO$_3$. (b) Total energy as a function of $c/a$ for 0.50BTBCO in the ferroelectric $P4mm$ phase. The crystal structure shows the octahedral environment of Co surrounded by O atoms rather than pyramidal environment due to reduction in $c$. The $A-$ and $B-$ sites are occupied by Bi/Ba(50\%+50\%) and Co/Ti(50\%+50\%), respectively.}
\label{fig3}
\end{figure}

The lattice parameters $a$ ($c$) of $x$BaTiO$_3-$(1-$x$)BiCoO$_{3}$ linearly increase (decreases) with increase the value of $x$ which results in a decrease in tetragonality ($c/a$) as shown in Figure~\ref{fig3}(a). The substitution of Ba$^{2+}$ for Bi$^{3+}$ and Ti$^{4+}$ for Co$^{3+}$ lead to these structural parameters changes. As the ionic radii of Ba$^{2+}$(1.56) ions are larger than the that of the Bi$^{3+}$(1.17), these cation substitutions induce isostatic pressure (chemical pressure) on the lattice. The reduction in tetragonality results in an octahedral coordination (as shown in Figure~\ref{fig3}(b)) rather than a pyramidal structure (as shown in Figure~\ref{fig2})for higher $x$ values in $x$BaTiO$_3-$(1-$x$)BiCoO$_{3}$ solid solution. This large lattice distortion is associated with reduction in the lone pair electron concentration coming from Bi$^{3+}$ ion and also the increase in $d^0$ ions interactions from Ti$^{4+}$ ion that gives covalency with increasing value of $x$.
\par
Our high resolution synchrotron diffraction data measurements made as a function of $x$ show that tetragonal to cubic phase transition occurs around $x$ = 0.5 composition (see Figure S1 of the supporting information). To verify this, we have performed energy vs. $c/a$ calculations for $x$BaTiO$_3-$(1-$x$)BiCoO$_{3}$ solid solution as a function of $x$. To perform this calculation, we have doubled the unit cell along $c-$axis to get a 1x1x2 supercell. If the crystal is cubic, one should get minimum energy for $c/a$ =2. However, the total energy vs $c/a$ curve showing minimum value at $c/a$ = 2.05 for the composition $x$ = 0.5 as evident from Figure~\ref{fig3}(b). It may be noted that small composition variations and high temperature synthesis processes can stabilize the paraelectric phase (i.e. a phase with $c/a$ =2) over the ferroelectric phase as our experimental results suggest. However drastic reduction of tetragonality with x indicate that the system is going towards the paraelectric phase with increase in $x$.

\subsection{Electronic and magnetic structure}
\begin{figure}[!t]
\centering
   \includegraphics[height=4cm]{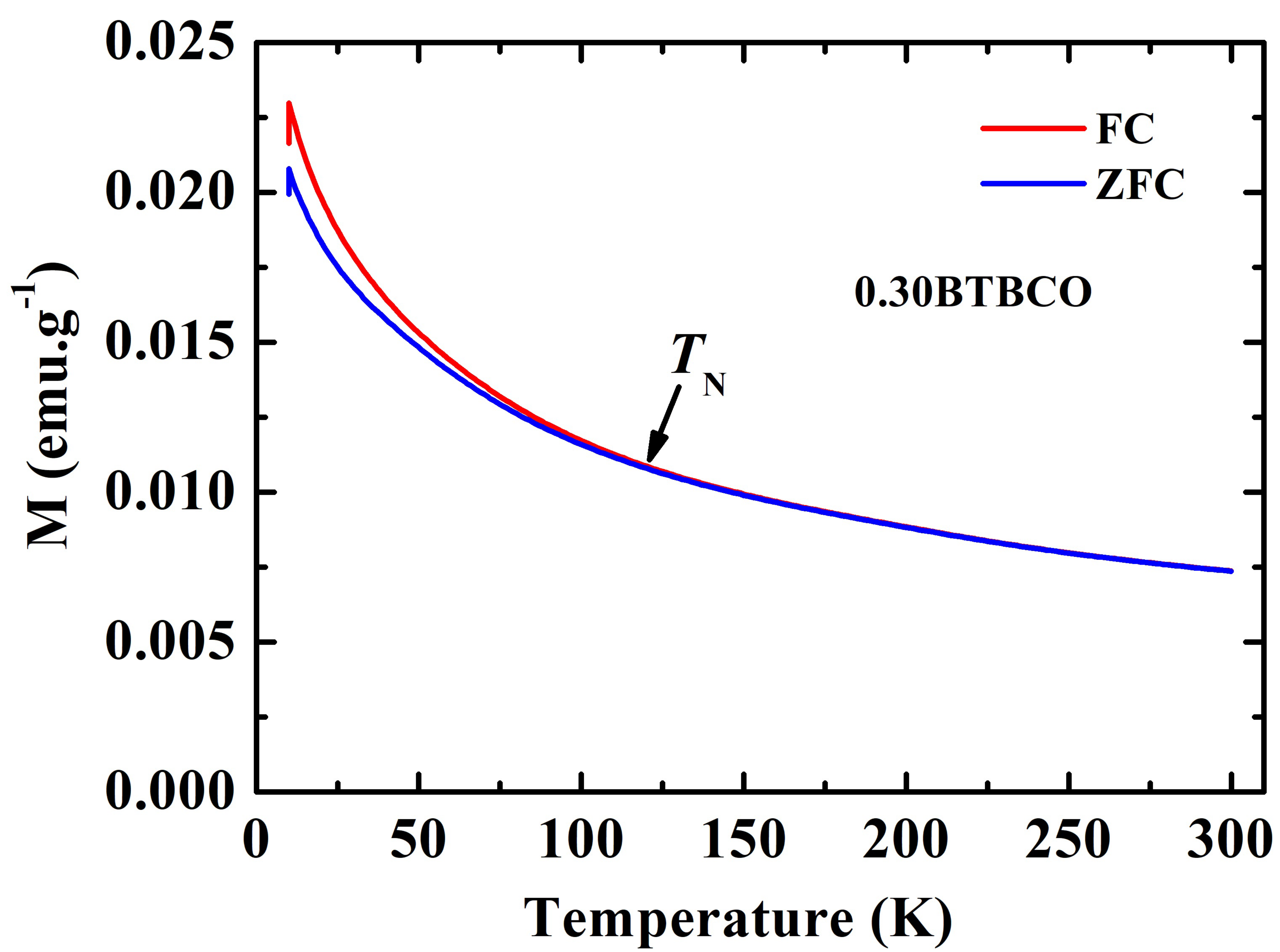}
   \hspace{0.10in}
    \includegraphics[height=4cm]{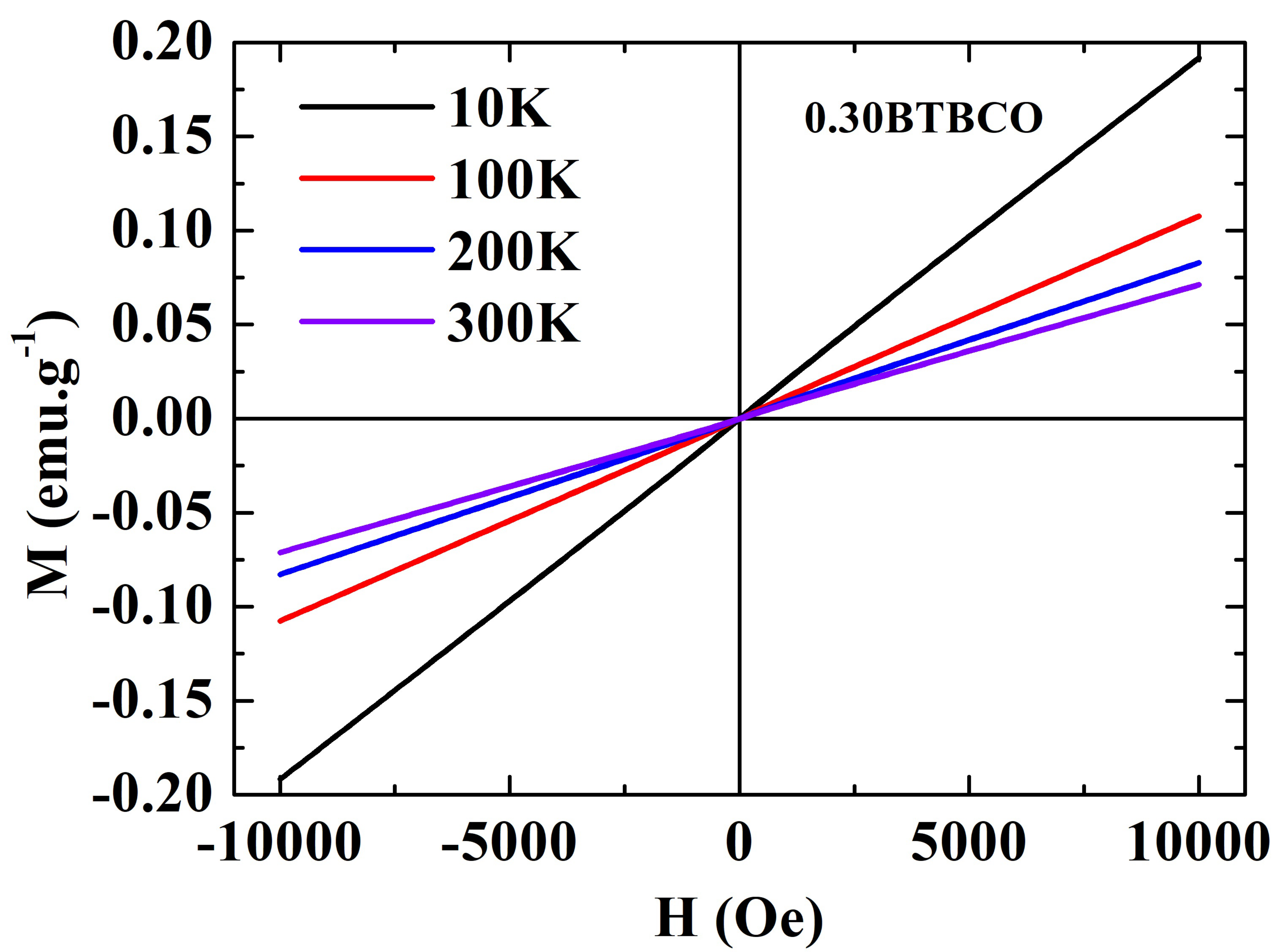}
   \hspace{0.10in}
\caption{\small(Color online) (a) ZFC/FC magnetization for 0.30BTBCO. (b) M$-$H loops taken at various temperatures for 0.30BTBCO}
\label{fig4} 
\end{figure}

As shown in Figure~\ref{fig4}(a), an anomaly was observed around 120\,K on the  zero field cooled (ZFC) measurement curve for 0.30BTBCO. Below the anomaly temperature, the field cooled (FC) curve deviates from the ZFC curve. Therefore, the anomaly indicates a phase transition from the paramagnetic to either an antiferromagnetic or a ferromagnetic phase. Since the \textit{x}BaTiO${_3}$-(1-\textit{x})BiCoO${_3}$ solid solution is a magnetically diluted system from BiCoO$_3$, it is expected to experience similar magnetic interaction as BiCoO$_3$, which is known to be antiferromagnetically ordered  below its N\'eel temperature $T_N\sim$ 470\,K. The N\'eel temperature for other 0.10BTBCO, 0.20BTBCO and 0.40BTBCO are mentioned in the Figure S2.(a), S2.(b) and S2.(c), respectively in the supporting information. Figure~\ref{fig4}(b) shows the magnetic field dependence of the magnetization (M$-$H curves), at different temperatures for 0.30BTBCO. It can be seen that the induced magnetization continuously rises (but does not saturate) with the increase of the applied magnetic field (until 10 kOe). The curves without any spontaneous magnetization indicates the antiferromagnetic nature of the sample.This is also a typical behavior of the well$-$known antiferromagnetic ordering in the BFPT system.~\cite{cotica2012ferroic} This magnetic ordering is originated by superexchange interactions in the Co$^{3+}-$O$-Co^{3+}$ arrangement. The detailed magnetic structure of 0.33BTBCO was determined by our total energy calculations. (See Figures S3(a), S3.(b) and S3.(c) for the M-H curve for 0.10BTBCO, 0.20BTBCO and 0.40BTBCO, respectively in the supporting information.)
 \par
\begin{figure}[!t]
\centering
\includegraphics[height=4cm]{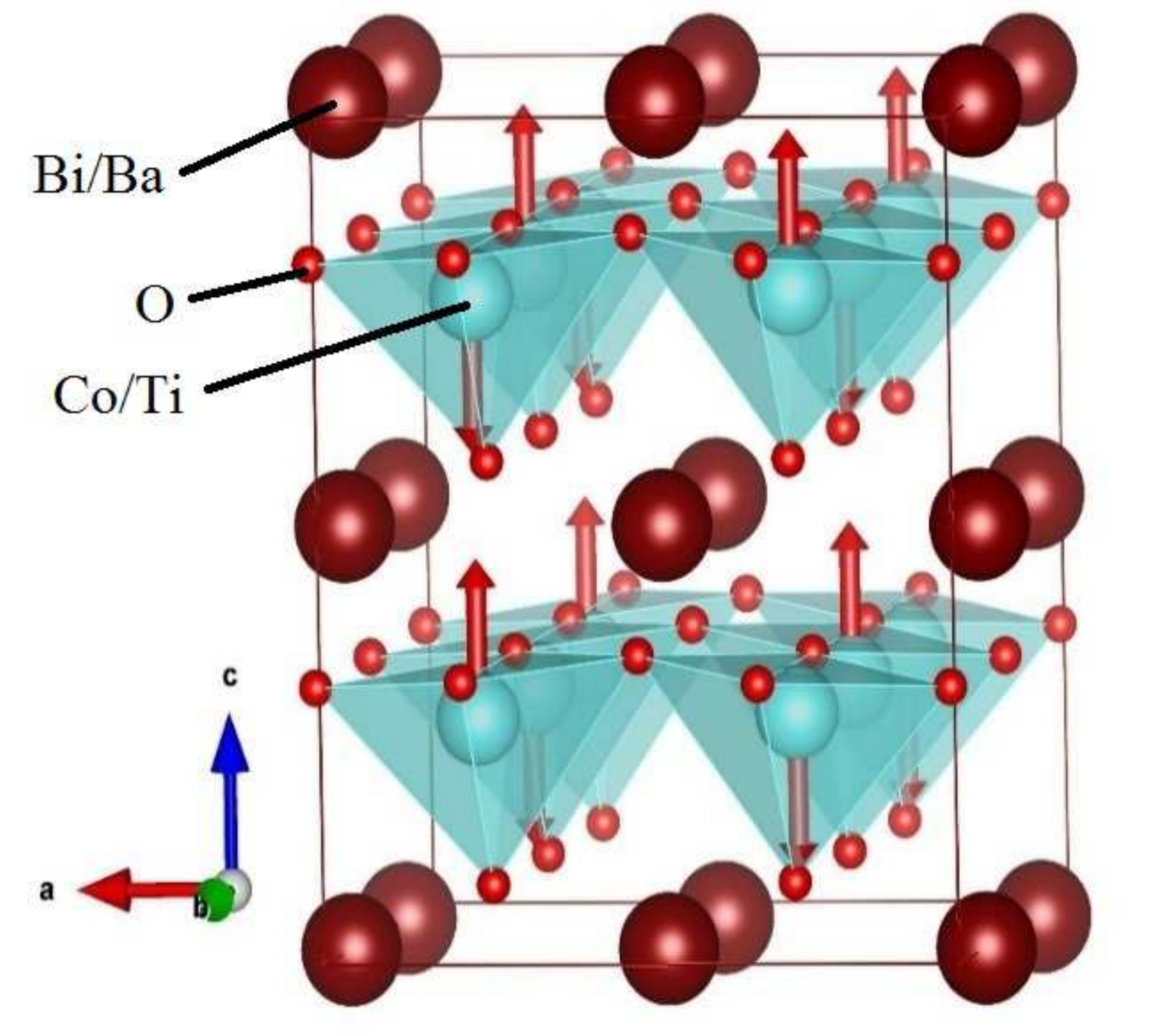}
  \hspace{0.10in}
\includegraphics[height=4cm]{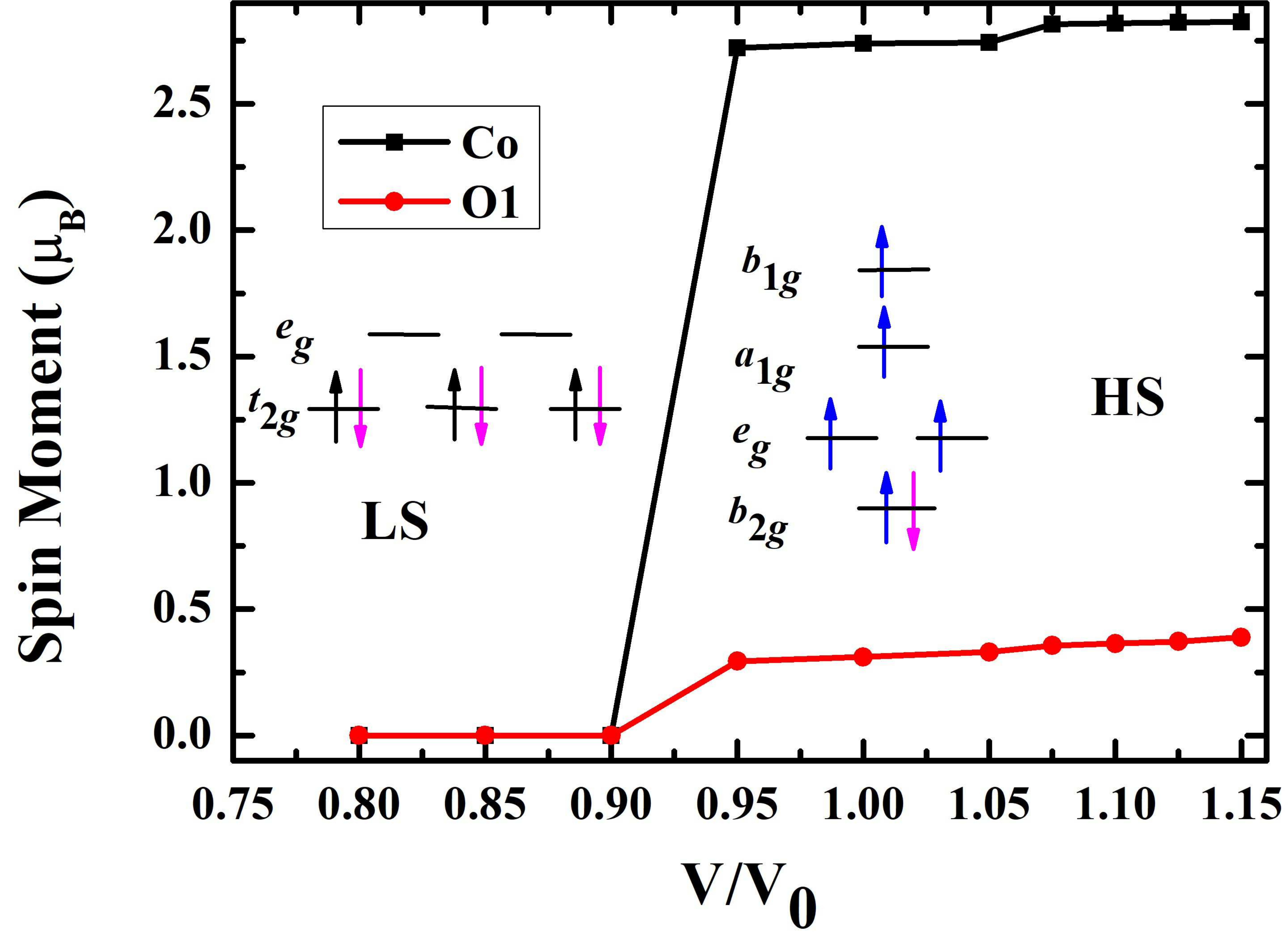}
   \hspace{0.10in}
\caption{\small(Color online) (a) Magnetic structure of 0.33BTBCO doubled along $c-$axis, where the arrows at the Co atoms indicate the $G-$type spin ordering. The square-pyramidal environment of Co/Ti atoms is indicated by polyhedra. The magnetic structure is found to be a tetragonal structure (space group $P4mm$) with lattice parameters $a$ = 7.4608 (7.4398),and c = 4.7897 (4.7196)\,\AA. (b) Variation of the spin magnetic moment Co and O1 (the spin moment at the O2 site is zero and not shown) with volume for tetragonal phase of 0.33BTBCO. The insets are schematic diagrams of the HS and LS spin configurations for the Co$^{3+}$ (3$d^6$) ions.}
\label{fig5}
\end{figure}
\begin{figure}[!t]
\centering
\includegraphics[scale=0.2]{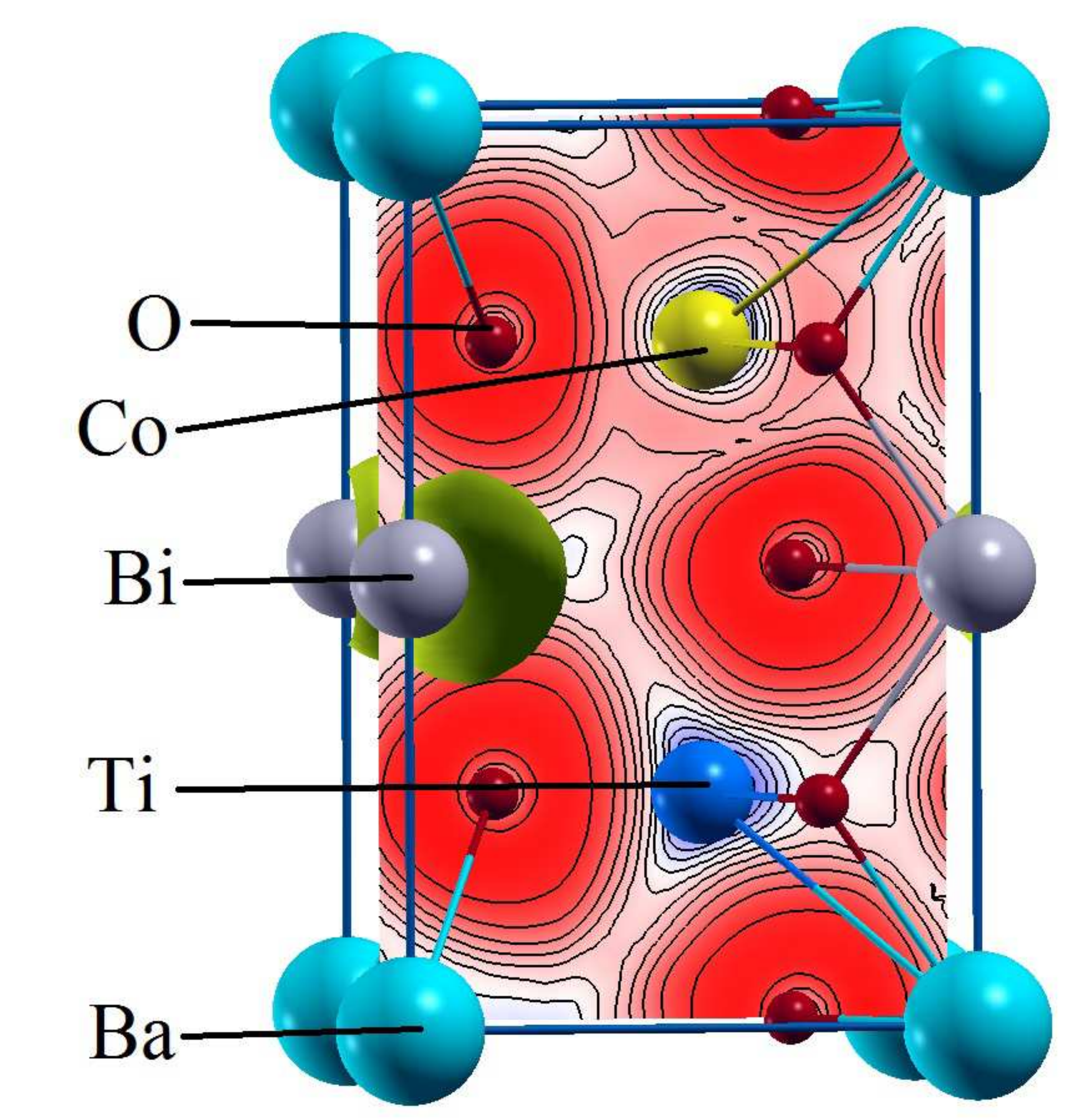}
\caption{\small(Color online)  Isosurface (at a value of 0.75) of the valence electron localization function of 0.5BTBCO in the ferroelectric $P4mm$ structure.}
\label{fig6}
\end{figure}
\begin{figure}[!t]
\centering
\includegraphics[scale=0.05]{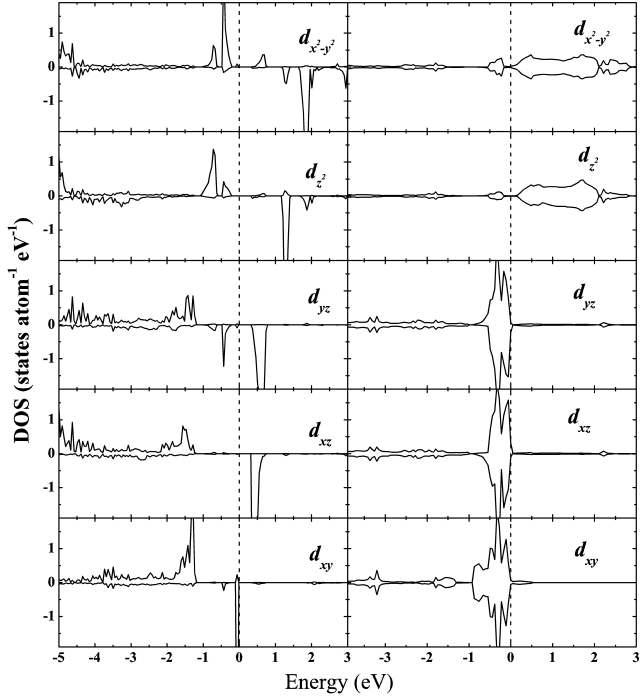}
\caption{\small Calculated orbital$-$projected density of states for 0.33BTBCO in the a) ferroelectric (left) and b) paraelectric (right) phases.}
\label{fig7}
\end{figure}

The relative total energies of various spin configurations with respect to the ground state configuration are summarized in Table.\ref{Table-1}. Our calculation shows that 0.33BTBCO is found to be a $G-$type antiferromagnet with both inter$-$ and intra$-$layer antiferromagnetic spin arrangements. The total energy of $G-$AFM is 39, 53 and 84\,meV lower than that of the $C-$AFM, $A-$AFM and F configurations, respectively. In order to understand the role of various magnetic configurations on the electrical properties of 0.33BTBCO, we have analyzed in detail the total and orbital projected density of states (DOS) for this system in different magnetic configurations. It can be seen from the optimized crystal structure that Co is surrounded by 5 oxygen atoms in a square pyramidal configuration (see Figure~\ref{fig5}(a)). In an ideal square pyramidal crystal field, Co 3$d$ splits into non-degenerate $b_{2g}$ ($d_{xy}$), doubly degenerate $e_g$ ($d_{xz}$, $d_{yz}$), non$-$degenerate $a_{1g}$($d_{z^2}$) and $b_{1g}$($d_{{x^2}-{y^2}}$). Generally, the $xy$ orbitals are higher in energy than that of the doubly degenerate $yz$ and $zx$ orbitals under a centrosymmetric tetragonal crystal field with $c/a >$ 1.\cite{cotton2008chemical}  
\begin{table}[!tbp]
\centering
\caption{The calculated total energies $\triangle$E (meV/f.u.) relative to the lowest energy states, the magnetic moments at Co site $\mu_{Co}$($\mu_B$), total magnetic moments $\mu_{Tot}$($\mu_B$), band gap E$_g$ (eV) values for different magnetic configurations for 0.33BTBCO}
\label{Table-1}
\setlength{\tabcolsep}{14pt}
\begin{tabular}{llllll}
\hline\hline
\textit{} & \textit{G-AF} & \textit{C-AF} & \textit{A-AF} & F     & NM    \\ \hline
$E$         & 0             & 39            & 53            & 84    & 120   \\
$\mu_{Co}$       & 2.588         & 2.537         & 2.779         & 2.818 & -     \\
$\mu_{Tot}$      & 3.049         & 2.949         & 3.429         & 3.563 & -     \\
E$_g$  & 0.35          & 0.32          & 0.27          & 0.19  & Metal \\ \hline\hline
\end{tabular}
\end{table}
It is the non$-$centrosymmetric displacement that modifies this energy arrangement among the $t_{2g}$ orbitals and renders the $xy$ orbital get occupied fully  with an insulating energy gap above it (see Figure~\ref{fig5}(b) inset). The presence of lone$-$pair electrons at the Bi site (seen from electron$-$localization$-$function (ELF) plot given in Figure~\ref{fig6}) and orbital$-$projected DOS for Co (see Figure~\ref{fig7}) indicate that the formal valence for Co ions can be assigned as 3+. In an ideal cubic perovskite$-$like lattice, the Co$^{3+}$ ions in the HS state prefer to form the $G-$AFM ordering due to the fact that the Pauli exclusion principle allows the transfer of an electron to the neighboring ion in an antiparallel direction only which is consistent with our results. The magnetic structure of a separate unit cell corresponds to the $G-$type AFM ordering is shown in Figure~\ref{fig5}(a) where each magnetic Co$^{3+}$ ion is surrounded by other four Co$^{3+}$ ions with spins directed opposite to that of the central ion. Figure~\ref{fig5}(b) shows the variation in the magnetic moment at the Co$-$site as a function of volume in the $G$-AFM state. From this curve it may be noted that when we compress the volume above 5\%, suddenly the magnetic moment at both Co and O1 site reduces to zero showing the magnetic to nonmagnetic transition in 0.33BTBCO. This effect is explained later with our fixed spin moment (FSM) plots. The calculated magnetic moments and the total energy with respect to the ground-state magnetic configuration for 0.33BTBCO are given in Table.\ref{Table-1} Like LaCoO$_3$, 0.33BTBCO also has Co ions in the 3+ states. So, similar to LaCoO$_3$ one could expect nonmagnetic solution for 0.33BTBCO. But, owing to the presence of lone pair electrons at the Bi sites, one could expect anisotropy in the exchange interaction also. Consistent with this expectation we found that the NM state is found to be 120\,meV/f.u. higher in energy than the $G-$AFM state in 0.33BTBCO. This suggests that the Bi lone pair electrons play an important role in not only the ferroelectric behaviors but also in the magnetic properties. 
\par
\begin{figure}[!t]
\centering
\includegraphics[scale=0.4]{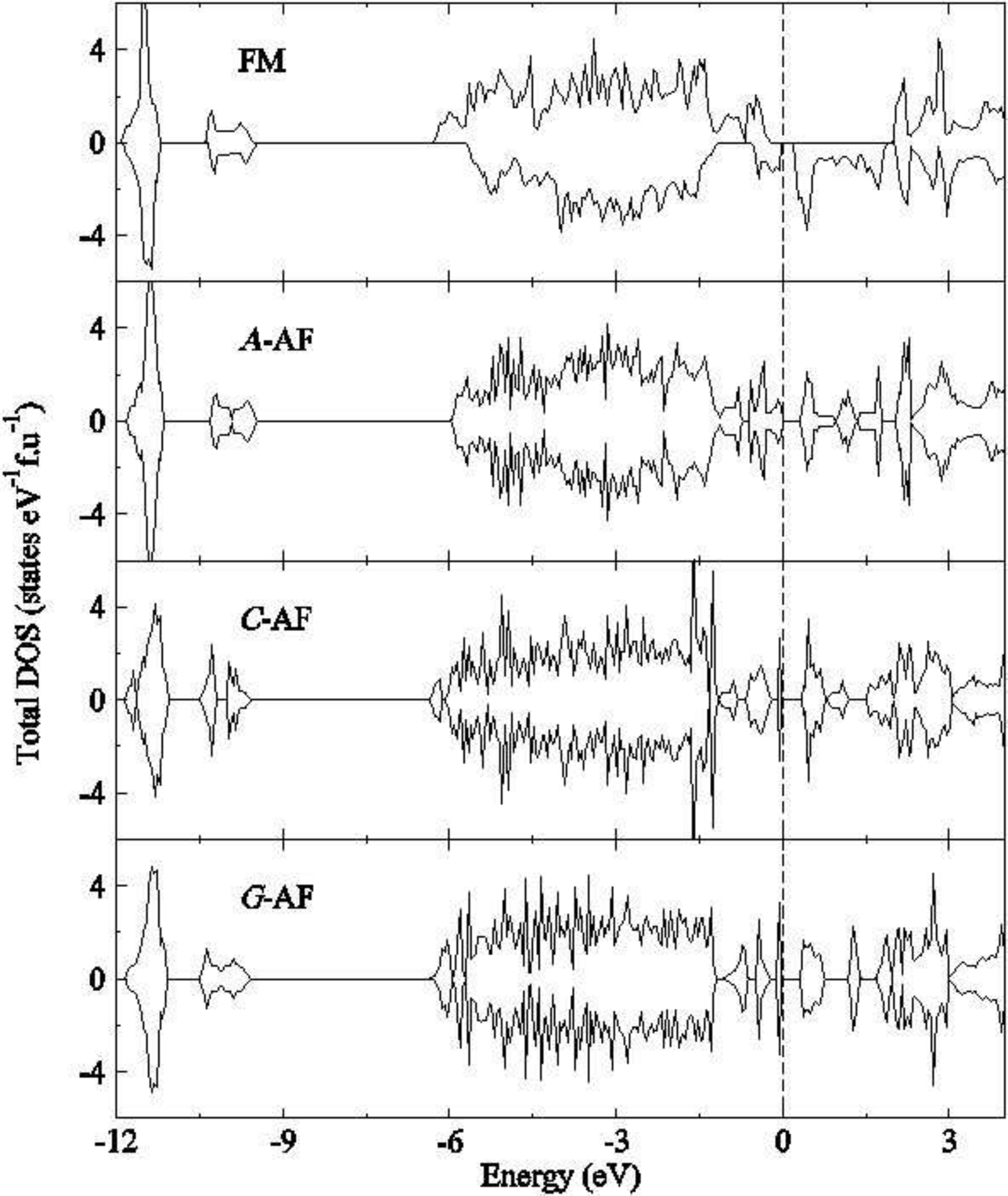}
\caption{\small Calculated total DOS for 0.33BTBCO in FM, $A-$AF, $C-$AF and $G-$AF configurations}
\label{fig8}
\end{figure}
\begin{figure}[!t]
\centering
\includegraphics[scale=0.07]{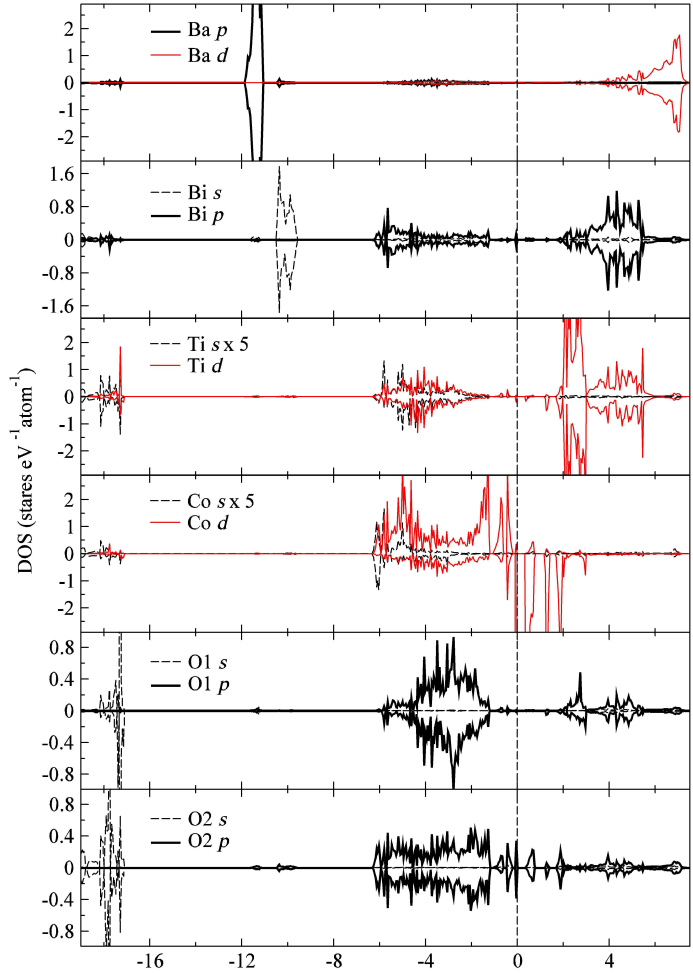}
\caption{\small(color online) Site and angular-momentum projected DOS for 0.33BTBCO for $G-$AFM configuration}
\label{fig9}
\end{figure}

The total DOS for different magnetic configurations of 0.33BTBCO are shown in Figure~\ref{fig8}. The total DOS for FM configuration shows semiconducting behavior with a very small band gap in the minority spin channel. When we introduce AFM ordering, the exchange interaction produces an exchange potential that effectively shifts the energy of Co 3$d$ band to lower energy and making the gap wider (see Table.\ref{Table-1}) as shown in Figure~\ref{fig8}. It is hereby found that the ground state not only depends upon the spin polarization but also depends on the magnetic ordering. From the partial DOS, a sharp peak can be seen around $-$11.5\,eV which is corresponding to Bi 6$s$ states.(see Figure~\ref{fig9}) In BaTiO$_3$,  the  Ti $d$ and  O $p$ strongly  hybridize making the charges redistribute between the two states, resulting in a fraction of Ti $d$ states to be occupied.\cite{kuroiwa2001evidence,filippetti2002coexistence} The Bi 6$p$, Co 3$d$, Ti 3$d$ and O 2$p$ states are distributed from $-$6\,eV to $E_F$ forming a broad band. This indicates the strong hybridization effect of Co 3$d-$O 2$p$, Bi 6$p-$O 2$p$ and Ti 3$d-$O 2$p$ states. The $p$ states of Ba atoms and the 2$s$ states for O1 and O2 are well-localized and are present around $-$11\,eV and $-$18\,eV, respectively. Hybridization between Co 3$d$ and the surrounding oxygen ligands is well known to lead to superexchange interactions in magnetic perovskites. 

\par

The hybridizations between occupied Bi 6$s$, Bi 6$p$ and O 2$p$ states are unexpected at the first sight. As these issues have great consequences on the properties of $x$BaTiO$_3-$(1-$x$)BiCoO$_{3}$, they can be understood as follows. The Bi 6$p-$O 2$p$ hybridization is evident from the PDOS plot shown in Figure~\ref{fig9} as they are energetically degenerate. The Bi lone pairs are expected to have primarily 6$s$ character. But according to Figure~\ref{fig6}, a lobe can be seen at the Bi site. To form a lobe, there must be some hybridization interaction between the neighbors. Watson and Parker\cite{watson1999ab,watson1999origin} suggested that it is due to the $p$ character from the anions that lone pairs are allowed to form lobe$-$shaped structures. If the lone pair 6$s$ electrons overlap with o $p$ states, it can attain a lobe shape structure. In order to have hybridization between Bi 6$s$ and O 2$p$ orbitals, they should be energetically degenerate. From Figure~\ref{fig9}, we can see that even though O 2$p$ states are primarily present at the top of the valence band, finite O $p$ electrons are present around $-$10\,eV where the Bi 6$s$ electrons are also present. So, O $p$ electrons participate in the hybridization interaction with the Bi 6$s$ electrons that bring nonspherical distribution of lone pair electrons.
\par
\begin{figure}[!t]
\centering
\includegraphics[scale=0.1]{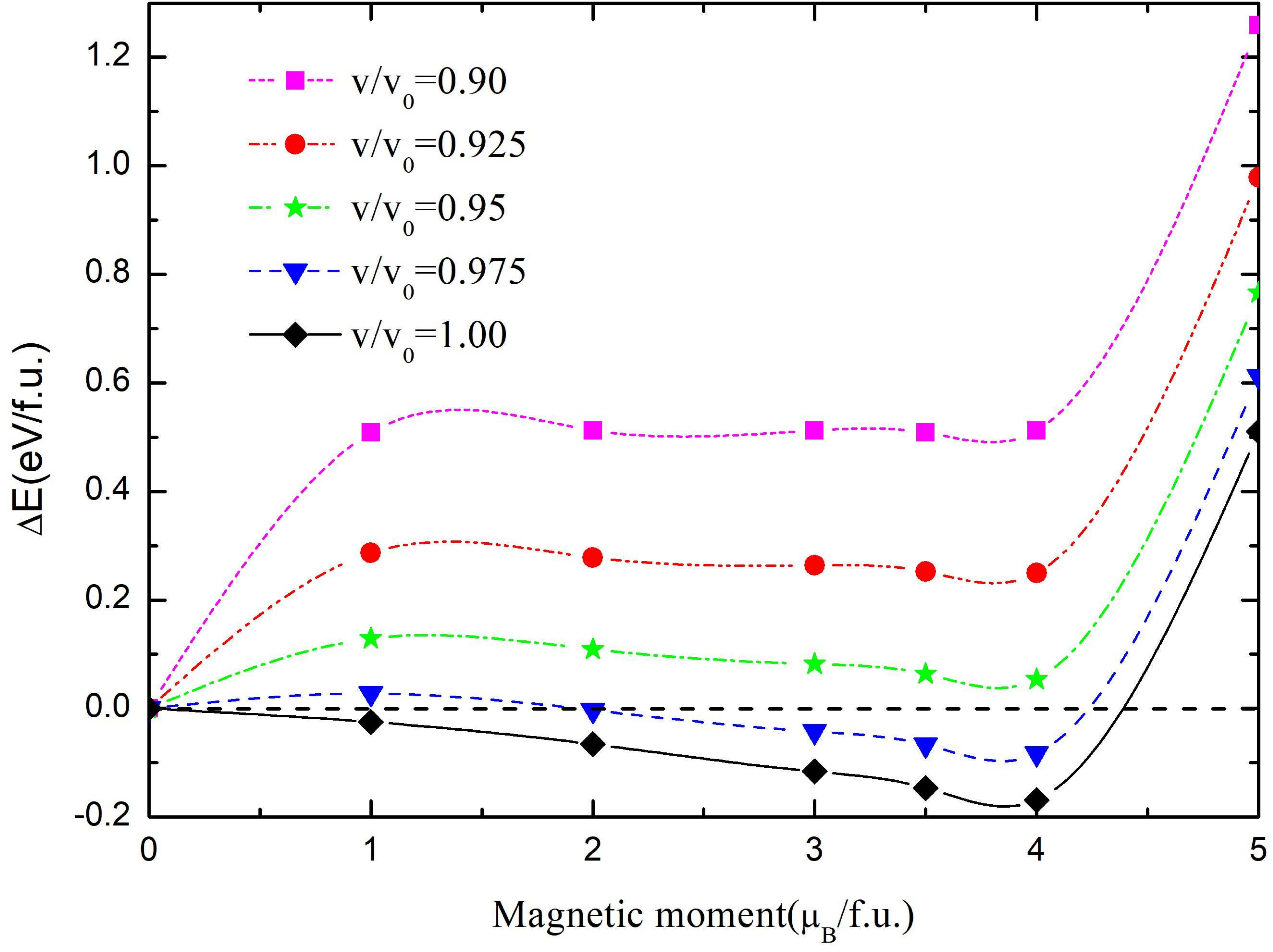}
\caption{\small Variation of total energy with magnetic moment for 0.33BTBCO obtained from the FSM calculations for different volumes. The total energy for non$-$spin polarized case is set to zero.}
\label{fig10}
\end{figure}

The calculated magnetic moment at the Co site varies between 2.537\,$\mu_B$ and 2.818\,$\mu_B$ per Co atom depending upon the magnetic configuration considered in our calculations (see Table.\ref{Table-1}). The calculated magnetic moments at the cobalt sites are not integer values, since the Co electrons have a hybridization interaction with the neighboring O ions. Because of this hybridization interaction we found that around 0.34\,$\mu_B$ magnetic moment is induced at each O site in the ferromagnetic state which are polarized along the same direction as that in the Co sites. As a result, the net total moment at in 0.33BTBCO is considerably larger (see the total moment in Table.\ref{Table-1}). The spin state of Co$^{3+}$ has been always a topic of interest for the scientific community from the last few decades due to exotic behavior arising from spin state transitions. It is generally believed that the Co$^{3+}$ ions adopt a low$-$spin (LS) state in octahedral CoO$_6$ coordination environments and the high$-$spin (HS) or intermediate spin (IS) state in pyramidal CoO$_5$ coordination environments. In pure ionic picture Co$^{3+}$ with six $d$ electrons will fill the energy levels either with a non-magnetic LS state ($b_{2g}^2$ $e_g^4$ $a_{1g}^0$ $b_{1g}^0$) or with a magnetic HS state ($b_{2g}^2$ $e_g^2$ $a_{1g}^1$ $b_{1g}^1$) having spin moments 0 and 4$\mu_B$, respectively. In the IS configuration ($b_{2g}^2$ $e_g^3$ $a_{1g}^1$ $b_{1g}^0$), one can expect a magnetic moment of 2\,$\mu_B$/Co. The electrons in transition metals are either participate in bonding or participate in magnetism. Hence the calculated magnetic moments are expected to be smaller than that for the pure ionic case owing to covalency effects. So, the calculated spin moment of 3.05$\mu_B$/f.u. in the ferroelectric phase may be interpreted as that for Co in HS state. The cooperative magnetism in 0.33BTBCO is originating from the partially filled and localized $b_{2g}^2$ $e_g^2$ $a_{1g}^1$ $b_{1g}^1$ orbitals in Co$^{3+}$ ions. Accordingly, our orbital$-$projected DOS in Figure~\ref{fig8} for the paraelectric and ferroelectric phases show LS and HS state for Co, respectively.  It is interesting to study the spin state transition as a function of volume  as shown in Figure~\ref{fig10}. These calculations were performed using FSM method with explicit supercell FSM for 33\% BaTiO$_3$ doped BiCoO$_3$ in which one uses magnetic moment $M$ as an external parameter and calculates the total energy as a function of $M$. The variation in total energy as a function of magnetic moment is shown for different unit cell volumes of 0.33BTBCO in Figure~\ref{fig10} as obtained from fixed spin calculations. The curves show that the energy difference between the HS and LS state is about 0.2\,eV/f.u. The HS state is found to be the ground state for equilibrium volume with moment around $\sim$3.6$\mu_B$ and around 5\% volume compression the LS state is found to be lower in energy than the HS state. This means that stong magnetovolume effect is present in this material and hence the HS to LS transition of Co$^{3+}$ also plays an important role in the giant volume contraction found from our experimental studies.

\par

\begin{figure}[!t]
\centering
\includegraphics[scale=0.2]{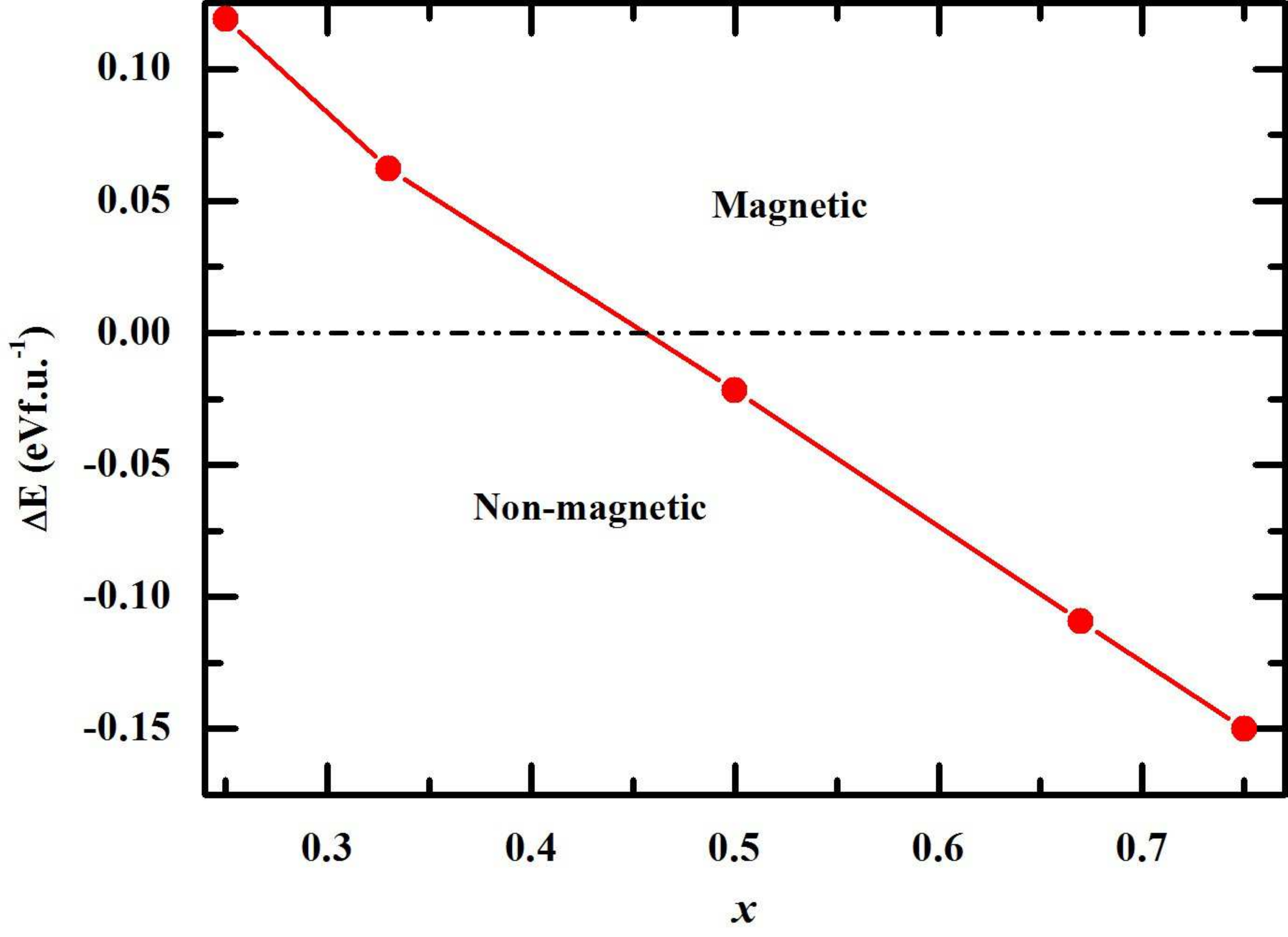}
\caption{\small Total energy difference between antiferromagnetic and nonmagnetic state of $x$BaTiO$_3-$(1-$x$)BiCoO$_{3}$ as a function of $x$.}
\label{fig11}
\end{figure}

From the total energy calculations, we have found that the ground state magnetic ordering for 0.33BTBCO and 0.25BTBCO is $G-$type AFM where both the inter$-$ and intra$-$planes are having antiferromagnetic ordering  as shown in fig \ref{fig5}(a).  For $x$ = 0.5, 0.67, and 0.75  the ground state emerges to be non$-$magnetic with Co$^{3+}$ ions in LS state.In order to understand the magnetic to nonmagnetic phase transition as a function of compositions (see Figure~\ref{fig11}), we have computed the total energy of $x$BaTiO$_3-$(1-$x$)BiCoO$_{3}$ in the tetragonal structure for the $G-$type antiferromagnetic as well as the nonmagnetic case as a function of $x$. Let us first analyze the possible electronic origin of the stabilization of the magnetic tetragonal phase for $x=$ 0.33 configurations. From the orbital$-$projected DOS given in Figure~\ref{fig7}, it can be seen that the nonbonding $t_{2g}$ electrons are piled up near Fermi level ($E_F$) in the cubic phase in the nonmagnetic configuration. This is an unfavorable condition for stability since the one-electron energy increases with increasing concentration of electrons closer to $E_F$. Hence a Peierls$-$Jahn$-$Teller$-$like instability\cite{peierls1991more} arises and the system stabilizes in the lower symmetric tetragonal structure. However, for tetragonal antiferromagnetic phase, the $t_{2g}$ levels split into doubly degenerate $a_{1g}$ state and singly occupied $b_{1g}$ state. As this crystal field splitting along with exchange splitting give a gain in the total energy of the system, the antiferromagnetic phase becomes lower in energy than the nonmagnetic phase (Table.\ref{Table-1}). Also it can be seen from Figure~\ref{fig8} that,  the $E_F$ falls on gap in the DOS in the tetragonal structure of 0.33BTBCO in the  $G-$AFM configuration. Since the location of $E_F$ on the gap gives extra contribution to the structural stability, 0.33BTBCO stabilizes in the G-type antiferromagnetic tetragonal structure. For $x=$ 0 and 0.25 also our calculations suggest that $x$BaTiO$_3-$(1-$x$)BiCoO$_{3}$ will stabilize in the $G-$type AFM structure with insulating behavior.  It is also clear from the composition vs total energy difference between nonmagnetic$-$magnetic configuration plot (see Figure~\ref{fig11}) that from magnetic to nonmagnetic transition is taking place around $x=$ 0.45. The syaytem is in the ferroelectric as well as magnetic in the composition range $x$ $<$ 0.45. So one could expect magnetoelectric behavior in the system. However, $x>$ 0.45, the system stabilizes in the nonmagnetic state with ferroelectric distortion.

\par

Two factors determine the magnetic properties of $x$BaTiO$_3-$(1-$x$)BiCoO$_{3}$ are the composition of the Co atoms and their distance with respect to O atoms. For higher $x$ values, the overlap interaction between the A -site cation with neighbours increases which will suppress the magnetic interaction. Further one increase the $x$, the amount of $d$ electrons in the system get reduced since Ti$^{4+}$ ion in principle don't have any $d$ electrons. This reduction in $d-$ electron also contribute to weakening of exchange interaction. As a result, the spontaneous magnetic ordering dissappers when go beyond $x=$ 0.45. Also for higher $x$ values not all Co atoms have enough Co atoms as neighbors to have exchange interaction which makes non-magnetic states stable at higher $x$ values. It may be noted that when we increase the BaTiO$_3$ content, the nearest distance between $B$-site also increases and hence one can expect enhancement in magnetic interaction due to localization of electrons. However, due to the increase in covalency and decrease in the d electron concentration brings nonmagnetic state stable over magnetic state for $x>$ 0.45. Moreover the tetragonality decreases with increase in $x$ and hence the interatomic distance between B$-$site cations increases (for a tetragonal lattice, the distance between $B-$site Cations is equal to the length of the $a-$axis) which weakens the super-exchange path and hinders the  antiferromagnetic interactions. 
\subsection{Born effective charge and spontaneous polarization}
In order to study the ferroelectric behaviour we have calculated the Born effective charges (BEC, $Z^*$) using the Berry phase approach generalized to spin-polarized systems.~\cite{resta1994macroscopic,king1993theory} These charges are important quantities in elucidating the physical understanding of piezoelectric and ferroelectric properties since they describe the coupling between lattice displacements and the electric field. Born effective charges are also indicators of long range Coulomb interactions whose competition with the short range forces leads to the ferroelectric transition. A large $Z^*$ value indicates that the force acting on a given ion due to the electric field generated by the atomic displacements is large even if the field is small, thus favoring a polarized ground state.  The calculated avarage diagonal components of $Z^*$ values for 0.33BTBCO are  $Z^*_{Bi}$= 5.39, $Z^*_{Ba}$ = 3.13, $Z^*_{Co}$= 3.29, $Z^*_{Ti}$=5.30, $Z^*_{O1}$ = -2.91, and $Z^*_{O2}$= -2.59.  It is known that the formal valence of Bi, Ba, Co, Ti and O in $x$BaTiO$_3-$(1-$x$)BiCoO$_3$ are +3, +2, +3, +4 and -2, respectively.  The value of $Z^*$ higher than formal oxidation state of an ion can reflect the presence of covalency in the bonding interaction of the ion with its neighbours.   We found that the $Z^*$ of the constituents in $x$BaTiO$_3-$(1-$x$)BiCoO$_{3}$ are significantly large revealing a large dynamic contribution superimposed to the static charge, that is, a strong covalency effect. This abnormally large BEC vaulue compared with formal oxidation state for the constituents is an important feature commonly also  found in other ferroelectric compounds.\cite{ghosez1995born,zhong1994giant} The maximum change for Co site BEC with respect to static charge is only ~10\%  indicating a relatively smaller covalency effect than that from Ti(+32\% higher than the formal valency). This is because the electrons in Co atoms contribute more to magnetism rather than bonding. This is in consistent with our conclusion from the  deduced magnetic moments at different atomic sites and also the bond length from the optimized structure of 0.33BTBCO (Co$-$O = 1.884\,\AA, Ti$-$O = 1.787\,\AA). 
\par
\begin{figure}[!t]
\centering
\includegraphics[scale=0.4]{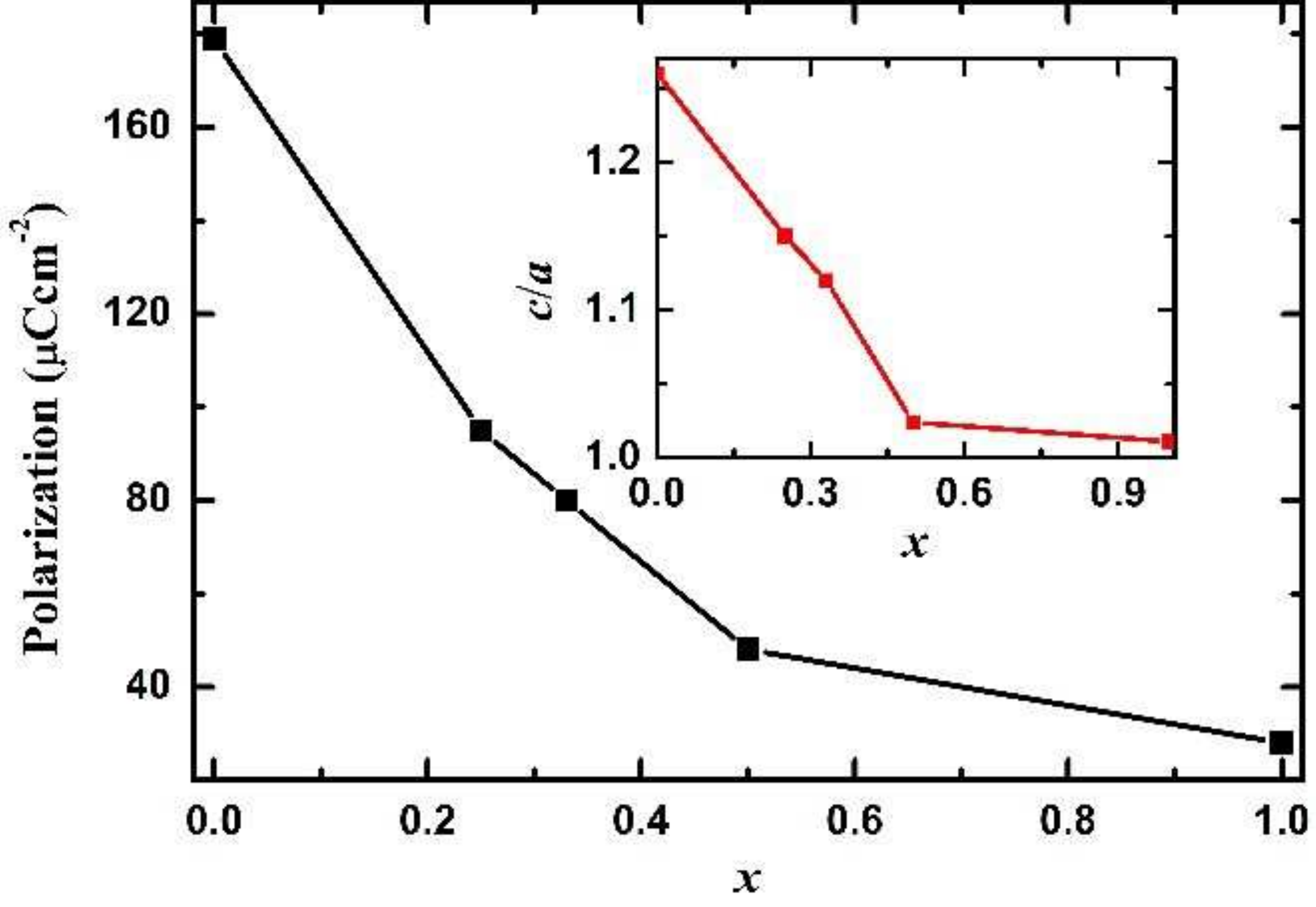}
\caption{\small Variation of calculated spontaneous polarization with $x$ for $x$BaTiO$_3-$(1-$x$)BiCoO$_3$. Inset shows the variation of $c/a$ with $x$}
\label{fig12}
\end{figure}

The Born effective charges can also be used to quantify the spontaneous polarization in $x$BaTiO$_{3}-$(1-$x$)BiCoO$_{3}$. As the  paraelectric phase produces zero polarization because of absence of dipole, the spontaneous polarization can be obtained by keeping the optimized paraelectric phase lattice parameter and changing the atomic position from paraelectric phase to ferroelectric phase. The introduction of BaTiO$_3$ leads to the decrease in length of $c-$axis and increase in length of $a-$axis, resulting the reduction of tetragonality ($c/a$). As a result, $x$BaTiO$_3-$(1-$x$)BiCoO$_3$ with higher value of $x$ produces an octahedral rather than a pyramidal coordination. The spontaneous polarization for 0.25BTBCO was calculated to be $\sim$90\,$\mu$Ccm$^{-2}$ which is comparable to that of multiferroic BiFeO$_3$\cite{ravindran2006theoretical} and much higher than typical piezoelectrics phase such as PbZr$_{0.52}$Ti$_{0.48}$O$_3$ (54\,$\mu$Ccm$^{-2}$)\cite{noheda2000tetragonal} and PbTiO$_3-$BiScO$_3$ (40\,$\mu$Ccm$^{-2}$).\cite{eitel2002preparation} Our partial polarization analysis shows that the polarization not only coming from the contribution of Bi$^{3+}$ lone pairs but also from the displacement of the ions when going from paraelectric to ferroelectric phase due to the strong Ba/Bi$-$O hybridization as well as the coupling interaction between cations such as Ti/Co and Ba/Bi.  Similar calculation using the nominal ionic charges yielded $P_s$ of $\sim$50 $\mu$Ccm$^{-2}$, almost 45\% less than the value obtained using the BECs. The reduction in the value of polarization obtained using nominal ionic charges is related to the covalency effect. The calculated polarization values are found to  decrease systematically  in $x$BaTiO$_3-$(1-$x$)BiCoO$_3$ with the increase in $x$ as shown in Figure~\ref{fig12}. This can be understood as follows. It is well known that the tetragonality has direct effect on electrical polarization of the tetragonal lattices. The introduction of BaTiO$_3$ leads to the decrease in $c-$axis and increase in $a-$axis, resulting the reduction of tetragonality ($c/a$). So there is a decrease in  trend is seen in the polarization value with increase in value of $x$. (see Figure S4 in the supporting information for polaraization vs $c/a$ plot). Also as we increase $x$, the Bi atoms are replaced by Ba atoms at the $A-$site resulting a reduction in the lone pair electron concentration in the lattice and consequently a decrease in polarization.
\section{Conclusion}
The present experimental and DFT calculation results indicate that $x$BaTiO$_3-$(1-$x$)BiCoO$_3$ possesses magnetoelectric behavior for $x$ < 0.45, which simultaneously show the coexistence of ferroelectricity and antiferromagnetism. Volume contraction has been observed in $x$BaTiO$_3-$(1-$x$)BiCoO$_3$ system at the ferroelectric$-$to$-$paraelectrcic phase transition point for BaTiO$_3$ concentration $<$ 0.45 where the Co$^{3+}$ makes an HS$-$LS transition. For higher $x$ values, it changes to nonmagnetic ferroelectric phase with reduction in spontaneous polarization value because of reducing tetragonality. We have reported that strong magnetoelectric coupling can be  accomplished in materials having a metamagnetism/magnetic instability. This finding opens up new possibility to find potential materials where one can vary the magnetic properties drastically (in extreme cases from a magnetic state to a nonmagnetic state, and vice versa as shown here) by the application of an electric field. Such strong coupling of magnetic and electric order parameters can be used for applications in spintronic devices. One may see data$-$storage media being alternatives to magneto$-$optical disks where the slow magnetic writing process is replaced by a fast magnetic inversion using electric fields. Thus we have described another route to giant magnetoelectricity, which is strongly related with an associated magnetic instability. This exciting concept will stimulate further research towards identifying novel multiferroics for practical applications. 

\section{Associated Content}
The Supporting Information is available free of charge on the ACS Publications website.
\section{Acknowledgments}
The authors are grateful to the Research Council of Norway for providing computing time at Norwegian supercomputer consortium (NOTUR). This research was supported by the Indo-Norwegian Cooperative Program (INCP) via Grant No. F. No. 58-12/2014(IC). L.P. wishes to thank Prof. Helmer Fjellv\r{a}g and Prof. Anja Olafsen Sj\r{a}stad for their fruitful discussions. L.P. also thanks Mr. Ashwin Kishore MR for critical reading of the manuscript.

\section*{References}

\bibliography{elsevier}

\end{document}